\documentclass[a4paper,11pt]{article}

\usepackage{graphicx}
\usepackage{natbib}

\newcommand{\titledpsp}[1]{\begin{center}{\large\textbf{\uppercase{#1}}}\end{center}}
\newcommand{\authordpsp}[1]{\begin{center}{\textbf{{#1}}}\end{center}}
\newcommand{\affildpsp}[1]{\begin{center}{\textit{{#1}}}\end{center}}

\begin{document}
% Do NOT change anything above this line

% This is the title of your abstract
\titledpsp{Properties of ion-cyclotron waves in the open solar corona}

% List here the authors and their affiliation as indicated
% using the sample format. Add/remove authors and affiliations
% as necessary.
\authordpsp{R. Mecheri}
\affildpsp{Centre de Recherche en Astronomie, Astrophysique et
Geophysique, CRAAG,
   Route de l'Observatoire, BP 63, Bouzareah 16340, Algiers, Algeria.}

\begin{abstract}

Remote observations of coronal holes have strongly implicated the
resonant interactions of ion-cyclotron waves with ions as a
principal mechanism for plasma heating and acceleration of the fast
solar wind. In order to study these waves, a WKB
(Wentzel-Kramers-Brillouin) linear perturbation analysis is used in
the work frame of the collisionless multi-fluid model where we
consider in addition to the protons a second ion component made of
alpha particles. We consider a non-uniform background plasma
describing a funnel region in the open coronal holes and we use the
ray tracing Hamiltonian type equations to compute the ray path of
the waves and the spatial variation of their properties.
\end{abstract}

\section{Introduction}\label{S-Introduction}

The ultraviolet spectroscopic observations made by SUMER (Solar
Ultraviolet Measurements of Emitted Radiation) and UVCS (Ultraviolet
Coronagraph Spectrometer) aboard SOHO indicated that heavy ions in
the coronal holes are very hot (hotter than protons by at least
their mass ratio, i.e. $T_{i}/T_{p}>m_{i}/m_{p}$) with high
temperature anisotropy \cite[see, e.g.,][]{Kohl,Wilhelm}. This
result is a strong indication for plasma heating through
ion-cyclotron resonance \citep[i.e., collisionless energy exchange
between ions and wave fluctuations, see][Chap~10]{Stix} involving
ion-cyclotron waves that are presumably generated in the lower
corona from small-scale reconnection events \citep{Axford95} or
locally through plasma micro-instabilities
\citep{Mecheri,Markovskii01} or turbulent cascade of low-frequency
MHD-type waves towards high-frequency ion-cyclotron waves
\citep{Li,Hollweg00,Ofman}. This notion led to renewed interest in
models involving ion heating by high-frequency ion-cyclotron waves
\citep{Isenberg,Hollweg00,Marsch,Vocks,Xie,Bourouaine}. For a
detailed review on resonant ion-cyclotron interactions in the
corona, see \citet{Hollweg02} and on plasma properties and physical
processes in coronal holes, see \citet{Cranmer09}. Therefore, to to
allow for a good physical interpretation of the observed data, it
became relevant to understand the properties of these waves in the
naturally multi-ion and highly non-uniform coronal plasma,
particularly in preparation to the future data which will be
provided by the recently approved high resolution ESA-NASA mission
"Solar Orbiter"
(planned launch in 2014).\\
Multi-ion linear mode analysis has been previously investigated by
\cite{Mann} in the frame work of the multi-fluid model but in the
simple case of a uniform background plasma. Our aim in this paper is
to go beyond this previous study and consider a realistic background
plasma with typical coronal profiles of density and temperature and
a typical 2D open magnetic field model describing a funnel region in
a coronal hole. We give a detailed study of ion-cyclotron waves by
performing a Fourier plane wave analysis using the collisionless
multi-fluid model where in addition to electrons (e) and protons
(p), we consider a second population of ions, namely alpha particles
(He$^{2+}$, indicated by $\alpha$) with a typical coronal abundance.
While neglecting the electron inertia, this model permits the
consideration of ion-cyclotron wave effects that are absent from the
one-fluid magnetohydrodynamics (MHD) model. Considering the WKB
(Wentzel-Kramers-Brillouin) approximation, we first solve locally
the dispersion relation and then perform a non-local wave analysis
using the ray-tracing theory. This theory allows to compute the ray
path of the waves as well as the spatial evolution of their
properties while propagating through the
funnel.\\
This paper is structured as follows. In Sect.~\ref{Background}, we
present the background plasma configuration we use in term of
density, temperature and the 2-D analytical funnel model describing
open magnetic field region in a coronal hole. Then in
Sect.~\ref{Basic}, we describe how the local and non-local
(ray-tracing) linear mode analysis are carried out using the
multi-fluid model. The results are presented and discussed in
Sect.~\ref{Numerical}, and finally our conclusions are given in
Sect.~\ref{Conclusion}.

\section{Background plasma configuration}\label{Background}

\begin{figure}
\begin{center}
$\begin{array} {c@{\hspace{-0.1in}}c} \hspace{-1.cm}
\includegraphics[width=7.cm]{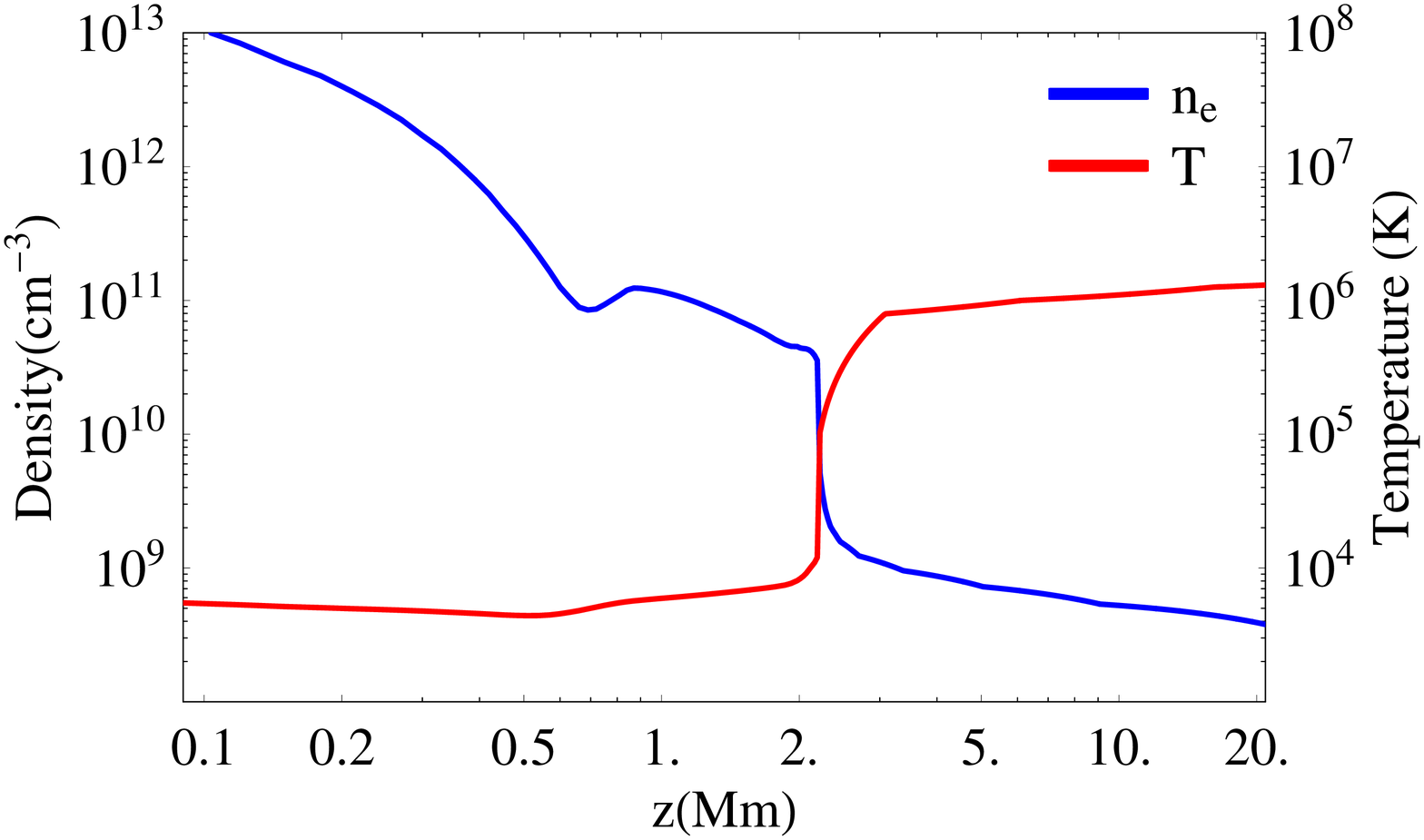}
\end{array}$
\hspace{-0.2cm} $\begin{array} {c@{\hspace{-0.1in}}c}
\includegraphics[width=5.1cm]{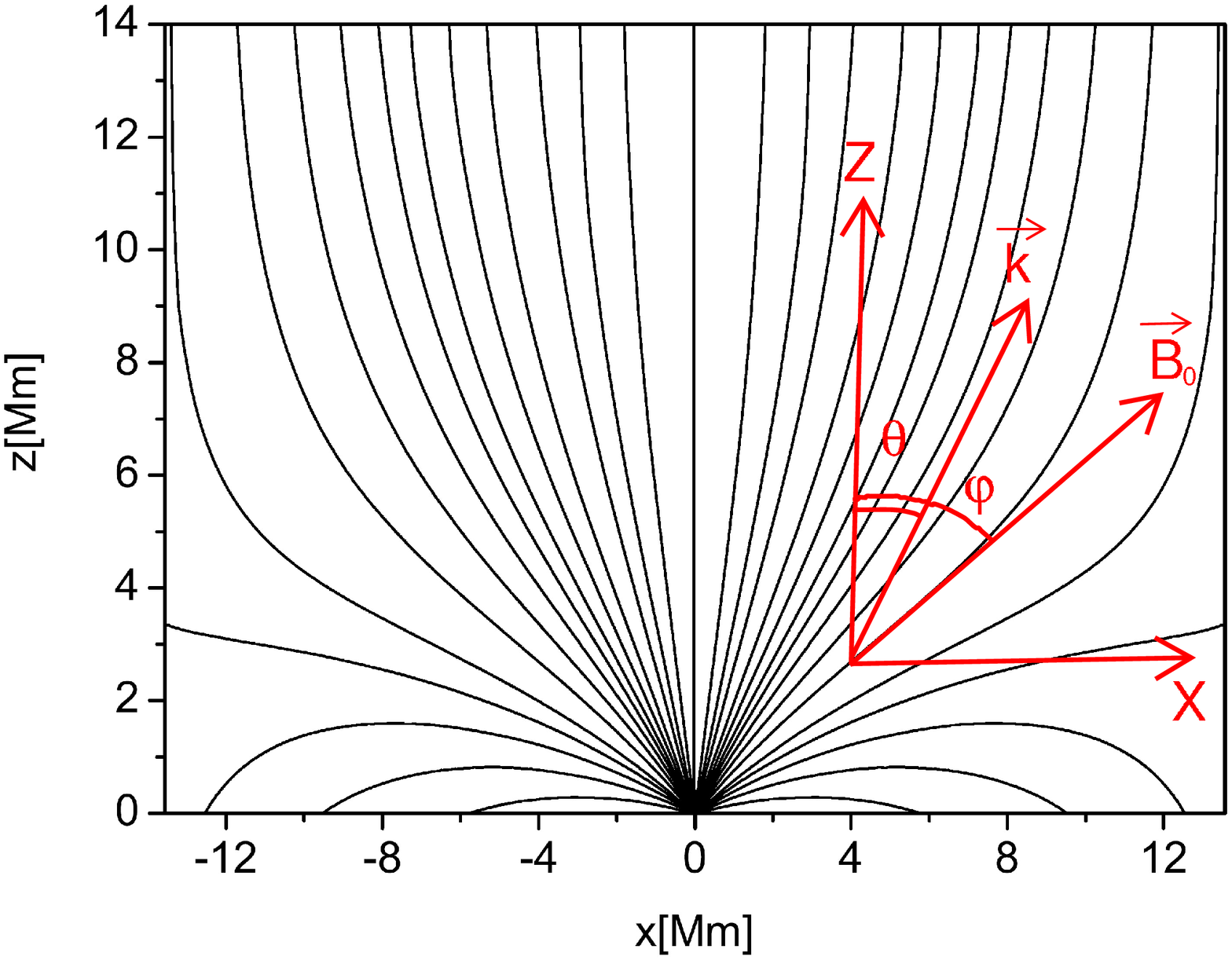}
\end{array}$
\end{center}
\vspace{-1cm} \caption{Left: Electronic density ($n_{e}$) and
temperature (\textit{T}) model-profiles of the chromosphere
\citep{Fontenla} and the lower corona \citep{Gabriel}. Right: Funnel
magnetic field geometry as obtained from a 2-D potential field model
\citep{Hackenberg}.} \label{fig1}
\end{figure}

As a background plasma density and temperature we use the model of
\citet{Fontenla} for the chromosphere and \citet{Gabriel} for the
lower corona (see Fig. \ref{fig1}). The 2-D potential-field model
derived by \citet{Hackenberg} is used to define the background
magnetic field (Fig. \ref{fig1}) representing a funnel.
Analytically, the two components of this model are:
\begin{eqnarray}
B_{0x}(x,z)&=&\frac{(B_{max}-B_{00})L}{2\pi(L-d)}ln\frac{cosh\frac{2\pi
z}{L}- cos(\frac{\pi d}{L}+\frac{2\pi x}{L})}{cosh\frac{2\pi z}{L}-
cos(\frac{\pi d}{L}-\frac{2\pi x}{L})}
\nonumber\\
B_{0z}(x,z)&=&B_{00}+(B_{max}-B_{00})\left[-\frac{d}{L-d}+\frac{L}{(L-d)\pi}\right.
\times\nonumber\\
&&\left(arctan\frac{cosh\frac{2\pi z}{L}~sin\frac{\pi d}{2L}+
sin(\frac{\pi d}{2L}+\frac{2\pi x}{L})}{sinh\frac{2\pi
z}{L}~cos\frac{\pi d}{2L}}+\right.
\nonumber\\
&&\left.\left.arctan\frac{cosh\frac{2\pi z}{L}~sin\frac{\pi d}{2L}+
sin(\frac{\pi d}{2L}-\frac{2\pi x}{L})}{sinh\frac{2\pi
z}{L}~cos\frac{\pi d}{2L}}\right)\right]
\end{eqnarray}
with the following typical relevant parameters: $L=30~ \textrm{Mm}$,
$d=0.34~\textrm{Mm}$, $B_{00}=11.8~\textrm{G}$, and
$B_{max}=1.5~\textrm{kG}$.

\section{Basic equations}\label{Basic}

The cold collisionless fluid equations for a particle species $j$
are:
\begin{equation}
  \frac{\partial n_{j}}{\partial
  t}+\nabla\cdot(n_{j}\textbf{v}_{j})=0,\label{mass}
\end{equation}
\vspace{-0.5cm}
\begin{equation}
  m_{j}n_{j}(\frac{\partial \textbf{v}_{j}}{\partial
  t}+\textbf{v}_{j}\cdot\nabla \textbf{v}_{j})-q_{j}n_{j}
  (\textbf{E}+\textbf{v}_{j}\times
  \textbf{B})=0,\label{momentum}
\end{equation}
\\where $m_{j}$, $n_{j}$, $q_{j}$ and $\textbf{v}_{j}$ are respectively the
mass, density, electric charge and velocity of a species $j$.
Subscript $j$ stands for electron $e$, proton $p$ or alpha particle
$\alpha$ (He$^{2+}$). The electric field $\textbf{E}$ and the
magnetic field $\textbf{B}$ are given by Faraday's law:
\begin{equation}
\nabla\times \textbf{E}=-\frac{\partial \textbf{B}}{\partial t}.
\label{field}
\end{equation}

\subsection{Linearization procedure}

The linear perturbation analysis is performed by expressing all the
quantities in the above equations as a sum of an unperturbed
stationary part (with subscript 0) and a perturbed part (with
subscript 1) much smaller than the stationary part:
\begin{eqnarray}
&&n_{j}=n_{0j}(z)+n_{1j},\textbf{B}=\textbf{B}_{0}(x,z)+\textbf{B}_{1},
\textbf{E}=\textbf{E}_{1},\textbf{v}_{j}=\textbf{v}_{1j}
\nonumber\\
&&\textrm{with:}~~n_{1j}\ll
n_{0j},\left|\textbf{B}_{1}\right|\ll\left|\textbf{B}_{0}\right|.
\label{perturbation}
\end{eqnarray}
Considering the quasi-neutrality (i.e., $\sum_{j}q_{j}n_{0j}=0$), no
ambient electric field (i.e., $\textbf{E}_{0}=0$) and no background
velocity (i.e., $\textbf{v}_{0j}=0$). The zero-order terms cancel
out when Eq. (\ref{perturbation}) is inserted into the Eqs.
(\ref{mass})-(\ref{field}). Neglecting the nonlinear products of the
first-order terms, we get a system of coupled linear equations:
\begin{equation}
i\omega\frac{n_{1j}}{n_{0j}} -i\textbf{k}\cdot\textbf{v}_{1j}=0,
\end{equation}
\vspace{-0.5cm}
\begin{equation}
i\omega\textbf{v}_{1j}+
\Omega_{j}\left(\frac{\textbf{E}_{1}}{\left|\textbf{B}_{0}\right|}
+\textbf{v}_{1j}\times\frac{\textbf{B}_{0}}{\left|\textbf{B}_{0}\right|}
\right)=\mathbf{0},
\end{equation}
\begin{equation}
i\textbf{k}\times \textbf{E}_{1}=i\omega \textbf{B}_{1},
\end{equation}
where all the perturbed quantities have been expressed in the form
of a Fourier plane wave. This Fourier analysis turns all derivatives
into algebraic factors, i.e. $\partial/\partial t\rightarrow -i
\omega$ and $\nabla\rightarrow i \textbf{k}$, where $\omega$ is the
wave frequency and $\textbf{k}$ the wave vector. The quantity
$\Omega_{j}=q_{j}B_{0}/m_{j}$ is the cyclotron frequency of species
$j$.

\subsection{Dispersion relation}

To derive the dispersion relation, the above linearized equations
have to be combined in order to obtain a linear relation between the
current density $\textbf{J}_{1}$ and the electric field
$\textbf{E}_{1}$:
\begin{equation}
\textbf{J}_{1}=\sigma\cdot\textbf{E}_{1},
\end{equation}
where $\sigma$ is the conductivity tensor which is related to the
dielectric tensor $\epsilon$ through the following relation:
\begin{equation}
\epsilon(\omega,\textbf{k},\textbf{r})=\textbf{I}+\frac{i}
{\omega\varepsilon_{0}}\sigma(\omega,\textbf{k},\textbf{r}).
\end{equation}
Finally, the local dispersion relation is obtained using the theory
of electrodynamics \citep[e.g.,][]{Stix}:
\begin{equation}
 D(\omega,\textbf{k},\textbf{r})=\textrm{Det}\left[\frac{c^{2}}{\omega^{2}}
 \textbf{k}\times(\textbf{k}\times\textbf{E})+\epsilon(\omega,
 \textbf{k},\textbf{r})\cdot\textbf{E}\right]=0,
\label{dispersion}
\end{equation}
where $c$ is the speed of light in vacuum and $\textbf{r}$ is the
large-scale position vector. We choose the wave vector $\textbf{k}$
to lie in the $x-z$ plane, with
$\textbf{k}=k(sin\theta,0,cos\theta)$.

\subsection{Principal waves properties}

In order to define certain wave properties we adopt the following
coordinate system: $\textbf{e}_{B_{0}}$ and $\textbf{e}_{k}$ are the
unit vectors respectively in the direction of the ambient field
$\textbf{B}_{0}$ and the wave vector $\textbf{k}$ (Fig. \ref{fig1}).
We assume here that these two unit vectors lie in the $x-z$ plane.
The unit vector perpendicular to $\textbf{e}_{k}$, but still in the
$x-z$, is $\textbf{e}_{\phi}$. The angle between $\textbf{B}_{0}$
and $\textbf{k}$ is $\phi$. In this configuration, the phase
velocity $\textbf{\textrm{v}}_{ph}$ and the group velocity
$\textbf{\textrm{v}}_{gr}$ are given by:
\begin{equation}
\textbf{v}_\mathrm{ph} = \frac{\omega}{k}~\textbf{e}_{k},~~~~~~
\textbf{v}_\mathrm{gr} =\nabla_{\textbf{k}}\omega = \frac{\partial
\omega}{\partial k}\textbf{e}_{k}+\frac{1}{k}\frac{\partial \omega}
{\partial \phi}\textbf{e}_{\phi}.
\end{equation}
The angle between the group velocity and the ambient field is given
by:
\begin{equation}
\psi=\textrm{arctan}\left[\left(\frac{\partial \omega}{\partial
k}sin\phi+\frac{1}{k}\frac{\partial \omega}{\partial
\phi}cos\phi\right)\bigg{/}\left(\frac{\partial \omega}{\partial
k}cos\phi-\frac{1}{k}\frac{\partial \omega}{\partial
\phi}sin\phi\right)\right].
\end{equation}
The helicity $\varrho$ (which represents the degree of circular
polarization) and the electrostatic part of the wave $\xi$ are given
by \citep{Krauss-varban}:
\begin{equation}
\varrho=-2\frac{\textrm{Re}(P_{\phi})}{1+|P_{\phi}|^{2}},
~~~~~~\xi=-\frac{1}{P_{\phi}}. \label{elec}
\end{equation}
where $P_{\phi}$ represents the $\phi$-component of the polarization
vector \cite[see, e.g.,][]{Shafranov}.

\subsection{Ray-tracing equations}

In the framework of the WKB approximation, the-ray tracing problem
consists at solving a system of ordinary differential equations of
the Hamiltonian form \citep{Weinberg}. These equations, which
represent the equations of motion for the wave frequency $\omega$,
the wave vector $\textbf{k}$, and the space coordinate $\textbf{r}$,
have been formulated by \citet{Bernstein}. In the simple case of a
Hermitian dielectric tensor, they are given by:
\begin{equation}
\frac{\textrm{d}\omega}{\textrm{d}t}=-\frac{\partial
D(\omega,\textbf{k} ,\textbf{r})/\partial t}{\partial
D(\omega,\textbf{k} ,\textbf{r})/\partial \omega}=0, \label{rt1}
\end{equation}
\vspace{-0.25cm}
\begin{equation}
\frac{\textrm{d}\textbf{k}}{\textrm{d}t}=~\frac{\partial
D(\omega,\textbf{k} ,\textbf{r})/\partial \textbf{r}}{\partial
D(\omega,\textbf{k} ,\textbf{r})/\partial \omega}, \label{rt2}
\end{equation}
\begin{equation}
\frac{\textrm{d}\textbf{r}}{\textrm{d}t}=-\frac{\partial
D(\omega,\textbf{k} ,\textbf{r})/\partial \textbf{k}}{\partial
D(\omega,\textbf{k} ,\textbf{r})/\partial \omega}. \label{rt3}
\end{equation}
Note that Eq.~(\ref{rt1}) can be set to zero, because the dispersion
relation does not explicitly depend on the time $t$ (i.e., the
background plasma is stationary). The above set of differential
equations represents an initial-value problem which can be solved
using initial conditions obtained from the local solutions of the
dispersion relation Eq.~(\ref{dispersion}).

\section{Numerical results}\label{Numerical}
\subsection{Local wave analysis}

\begin{figure}
\begin{center}
$\begin{array} {c@{\hspace{0.in}}c@{\hspace{0in}}c}
\multicolumn{1}{c}{\hspace{1.2cm}\mbox{\small Two-fluid (e-p)}} &
\multicolumn{1}{c}{\hspace{3.3cm}\mbox{\small Three-fluid
(e-p-$\textrm{He}^{2+}$)}}
\end{array}$
\includegraphics[width=11.8cm]{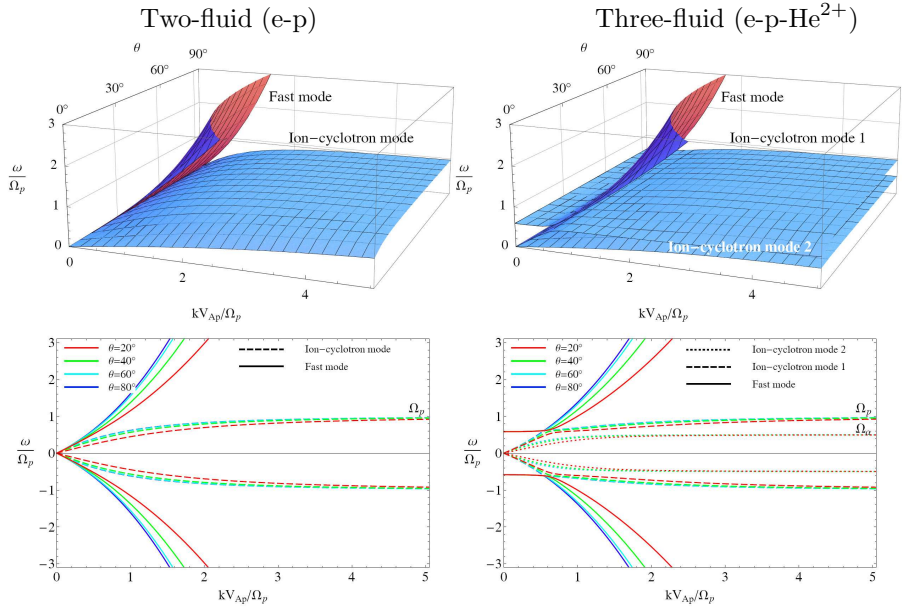}
\end{center}
\vspace{-0.3cm} \caption{Top panel: Two- and three-fluid dispersion
surfaces at the location (x=7.5 Mm, z=2.5 Mm) in the coronal funnel
characterized by an inclination angle $\varphi\approx79^{\circ}$ of
the magnetic field with respect to the $z$-direction (see Fig.
\ref{fig1}). Here $\omega$ and $k$ are normalized, respectively, to
the proton cyclotron frequency $\Omega_{p}$ and the inertial length
$\Omega_{p}/V_{Ap}$, with $V_{Ap}$ representing the proton
Alfv\'{e}n speed. Bottom panel: Two- and three-fluid dispersion
curves at the same location and for different propagation angles
$\theta$ with respect to the $z$-direction.} \label{fig2}
\end{figure}
\begin{figure}
\begin{center}
\includegraphics[width=12cm]{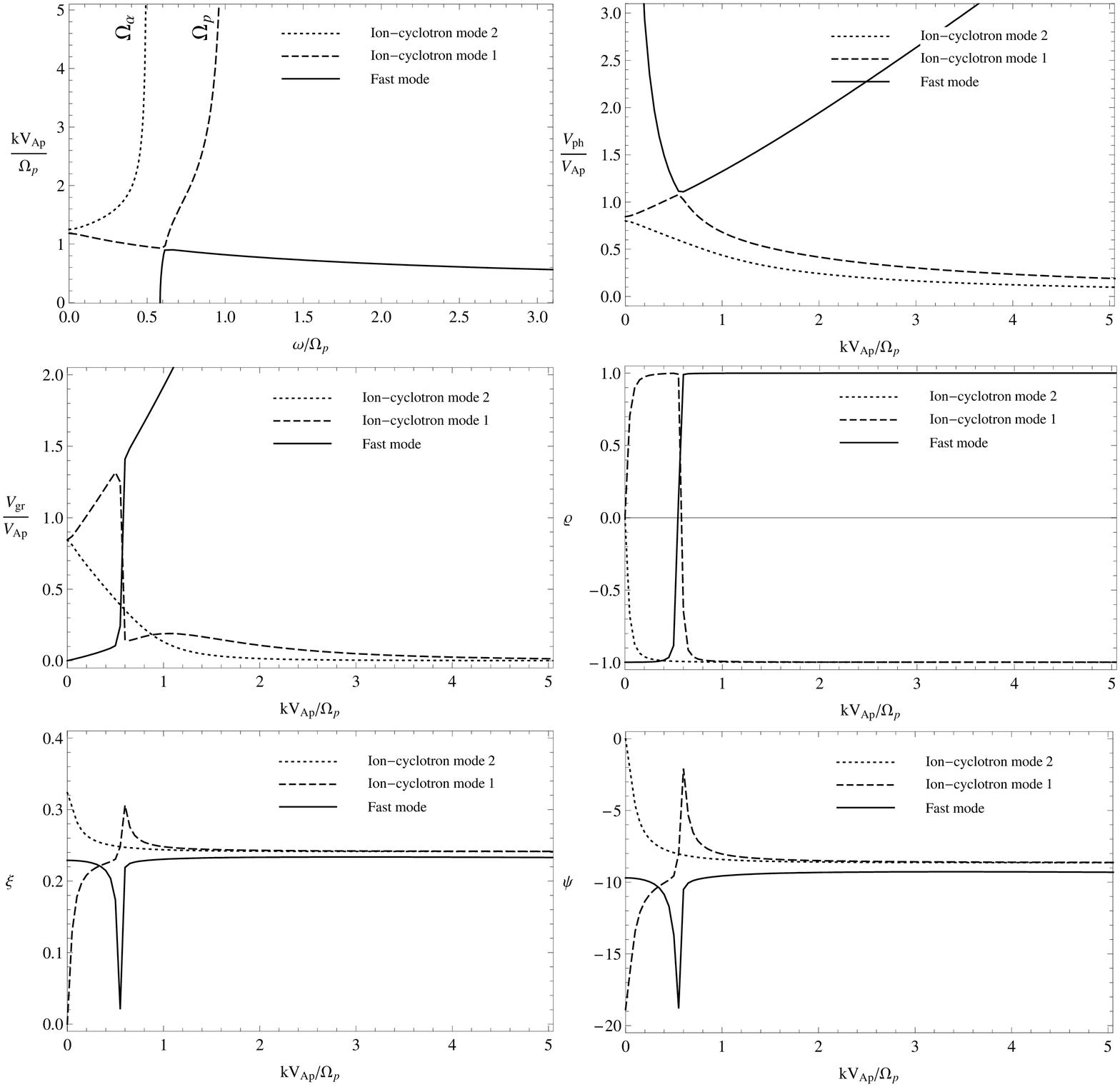}
\end{center}
\vspace{-0.3cm} \caption{Basic wave properties at the location
(x=7.5 Mm, z=2.5 Mm) in the coronal funnel characterized by an
inclination angle $\varphi\approx79^{\circ}$ of the magnetic field
with respect to the $z$-direction (see Fig. \ref{fig1}). The same
normalization as in Fig. (\ref{fig2}) is used. These results are
given for the case of a propagation angle $\theta=50^{\circ}$ with
respect to the $z$-direction. The wave properties are : \textbf{(a)}
refractive index $k\textrm{V}_{Ap}/\omega$, \textbf{(b)} phase
velocity V$_{ph}$, \textbf{(c)} The group velocity V$_{gr}$,
\textbf{(d)} helicity (degree of circular polarization) $\varrho$,
\textbf{(f)} electrostatic part of the wave $\xi$, \textbf{(e)}
angle between the direction of the group velocity and the ambient
magnetic field $\psi$.}\label{fig3}
\end{figure}

We consider a three-fluid cold plasma made of electron (e), protons
(p) and alpha particles (He$^{2+}$, indicated by $\alpha$) with a
density $n_{\alpha}=0.1n_{p}$. The dispersion relation
Eq.(\ref{dispersion}) is solved numerically at the funnel location
(x=7.5 Mm, z=2.5 Mm) which correspond to a limit region between open
and closed magnetic field lines where ion-cyclotron could
potentially be generated from small scales reconnection events
\citep{Axford95}. This location is characterized by a magnetic field
inclination angle of $\varphi\approx79^{\circ}$ with respect to the
normal on the solar surface. For comparison purpose, the dispersion
diagrams of a two-fluid (e-p) cold plasma are also presented on
Fig.(\ref{fig2}). In this case, Eq.(\ref{dispersion}) is a quadratic
polynomial of degree 4, which means that 2 modes exist and each one
is represented by an oppositely propagating ($\omega>0$ and
$\omega<0$) pair of waves. These modes, for which the dispersion
curves and surfaces are shown on the left panels of Fig.
(\ref{fig1}), are the Ion-Cyclotron mode (IC) and the Fast mode.
They are respectively the extensions of the usual Alfv\'{e}n and
Fast MHD modes into the high frequency domain around
$\omega\approx\Omega_{p}$ where they are dispersive as compared to
their non-dispersive MHD character at very low frequency
($\omega\ll\Omega_{p}$). In the case of the three-fluid model,
Eq.(\ref{dispersion}) is a quadratic polynomial of degree 6. Thus, 3
modes exist which in term of phase velocity ordering are: the
Ion-Cyclotron mode 1, the Ion-Cyclotron mode 2 and the Fast mode.
Indeed, the results represented on the right panels of
Fig.\ref{fig2} reveals the appearance of a second ion-cyclotron mode
IC2 in addition to IC1 mode and the fast mode already present in the
two-fluid model. For large k, the two IC modes reach a resonance
regime at each one of the cyclotron frequencies, i.e.
$\omega=\Omega_{p}$ and $\omega=\Omega_{\alpha}=\Omega_{p}/2$. The
heavy ion component strongly influences the dispersion branches at
small frequency $\omega/\Omega_{p}<1$ and small wave number
$\textrm{kV}_{Ap}/\Omega_{p}<1$ ($\textrm{V}_{Ap}$ is the proton
Alfv\'{e}n speed). Indeed, we note the appearance of a Fast mode
cut-off frequency approximately at
$\omega_{co}/\Omega_{p}\approx0.583$ in agreement with the algebraic
relation given by \cite{Melrose}:
\begin{equation}
\omega_{co}\approx
\Omega_{p}\frac{Z_{\alpha}}{A_{\alpha}}\left[\frac{1+A_{\alpha}
(n_{0\alpha}/n_{0p})}{1+Z_{\alpha}(n_{0\alpha}/n_{0p})}\right],
\end{equation}
with $A_{\alpha}=m_{\alpha}/m_{p}=4$ and the charge number
$Z_{\alpha}=2$. %resulting to $\omega_{co}/\Omega_{p}\approx0.583$.
Also, at k$\textrm{V}_{Ap}/\Omega_{p}\approx$0.5, the fast mode and
the IC2 mode couple and convert at a frequency corresponding to the
so-called cross-over frequency $\omega_{cr}/\Omega_{p}\approx0.612$,
also in conformity which the relation derived by \cite{Hollweg02}:
\begin{equation}
\omega_{cr}\approx \Omega_{p}
\left[\frac{Z_{\alpha}(n_{\alpha}/n_{p}+Z_{\alpha}/A_{\alpha}^{2})}
{1+Z_{\alpha}n_{\alpha}/n_{p}}\right]^{1/2},
\end{equation}
On Fig.(\ref{fig3}), we notice that the two IC modes starts at very
low frequency $\omega/\Omega_{p}\ll 1$ with a linear polarization
($\varrho=0$) in accordance with MHD theory. For increasing k, they
become very rapidly circularly polarized, right handed
($\varrho=+1$) for the IC1 mode and left handed ($\varrho=-1$) for
the IC2 mode. However, due to the coupling with the Fast mode, the
IC1 mode transforms rapidly into a left-handed circularly-polarized
mode at large k. At large k, these two modes reach a resonance
regime at each one of the ion-cyclotron frequency
$\omega=\Omega_{p}$ and $\omega=\Omega_{\alpha}=\Omega_{p}/2$,
regime which is characterized by an infinite refractive indices
$\textrm{kV}_{Ap}/\Omega_{p}$ (see Fig. \ref{fig3}). At very small
k, the IC2 mode is characterized by a group velocity mainly parallel
to the ambient field (i.e., $\psi\approx0$) but slightly deviates
from it $\mid\psi\mid\approx8^{\circ}$ for large k (see Fig.
\ref{fig3}). Since the direction of the group velocity indicates the
flow direction of the wave energy, this means that the IC2 mode
transport its energy mainly along the background magnetic field.
This mode is partially electrostatic with $0.23<\xi<0.33$. On the
other hand, at small k the IC1 mode is quasi-electrostatic (i.e.,
$\xi\approx0$) with an energy flow fairly oblique to the magnetic
field with $\mid\psi\mid\approx 20^{\circ}$ decreasing however to
$\mid\psi\mid\approx 8^{\circ}$ for large k. The Fast mode starts at
the cut-off frequency $\omega_{co}/\Omega_{p}=0.583$ at
$\textrm{k}=0$ and is initially left-hand polarized ($\varrho=-1$).
Owing to mode coupling with the IC1 mode, it rapidly changes its
sense of polarization around $\textrm{kV}_{Ap}/\Omega_{p}=0.5$ to
become a completely right-hand circularly-polarized wave
($\varrho=+1$). The fast mode is weakly electrostatic $\xi<0.23$ and
its group velocity propagates rather obliquely to the ambient
magnetic field ($10^{\circ}<\mid\psi\mid<19^{\circ}$).

\subsection{Non-local wave analysis: Ray tracing}

\begin{figure}
\begin{center}
$\begin{array} {c@{\hspace{0.in}}c@{\hspace{0in}}c}
\multicolumn{1}{c}{\hspace{-0.5cm}\mbox{\small Ion-Cyclotron mode
2}} & \multicolumn{1}{c}{\hspace{0.8cm}\mbox{\small Ion-Cyclotron
mode 1}} & \multicolumn{1}{c}{\hspace{1.5cm}\mbox{\small Fast mode}}
\end{array}$
\includegraphics[width=12.cm]{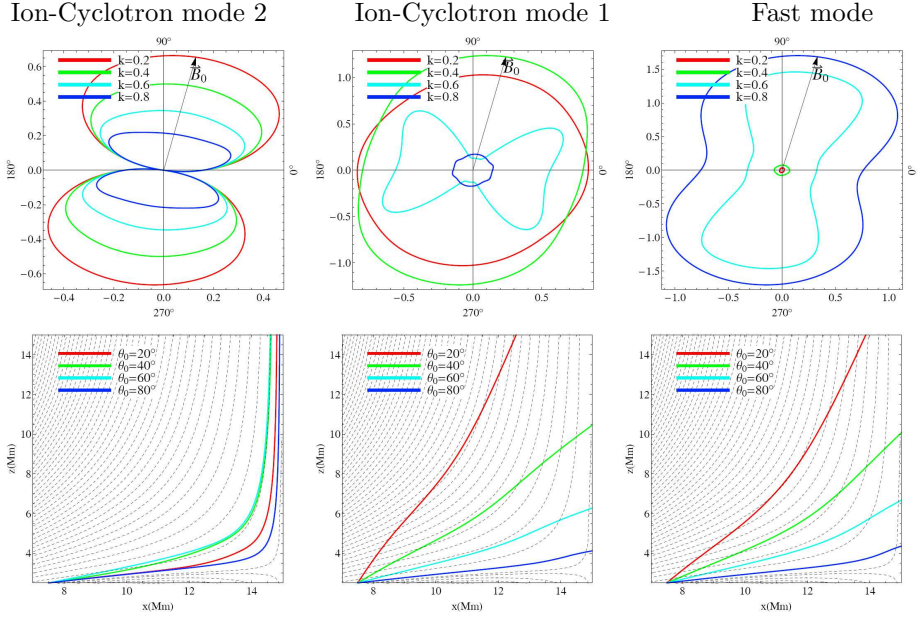}
\end{center}
\vspace{-0.3cm} \caption{Top panels: Topology of the group velocity
of the modes for different normalized wave number k and at the
location x=7.5~Mm,~z=2.5~Mm, with magnetic field $\textbf{B}_{0}$
inclination-angle $\varphi\approx79^{\circ}$. The length of the
radius at an angle of propagation $\theta$ counted from the
horizontal axis ($\theta=0^{\circ}$) is equal to the group velocity.
Bottom panels: The ray trajectory of the modes launched at the
location (x$_{0}$=7.5 Mm, z$_{0}$=2.5 Mm) in the coronal funnel,
with initial normalized wave number k$_{0}$=0.2 and at different
initial angles of propagation
$\theta_{0}=20^{\circ},40^{\circ},60^{\circ}~\textrm{and}~80^{\circ}$.}
\label{fig4}
\end{figure}
\begin{figure}
\begin{center}
$\begin{array} {c@{\hspace{0.in}}c@{\hspace{0in}}c}
\multicolumn{1}{c}{\hspace{-0.3cm}\mbox{\small Ion-Cyclotron mode
2}} & \multicolumn{1}{c}{\hspace{0.8cm}\mbox{\small Ion-Cyclotron
mode 1}} & \multicolumn{1}{c}{\hspace{1.5cm}\mbox{\small Fast mode}}
\end{array}$
\includegraphics[width=12cm]{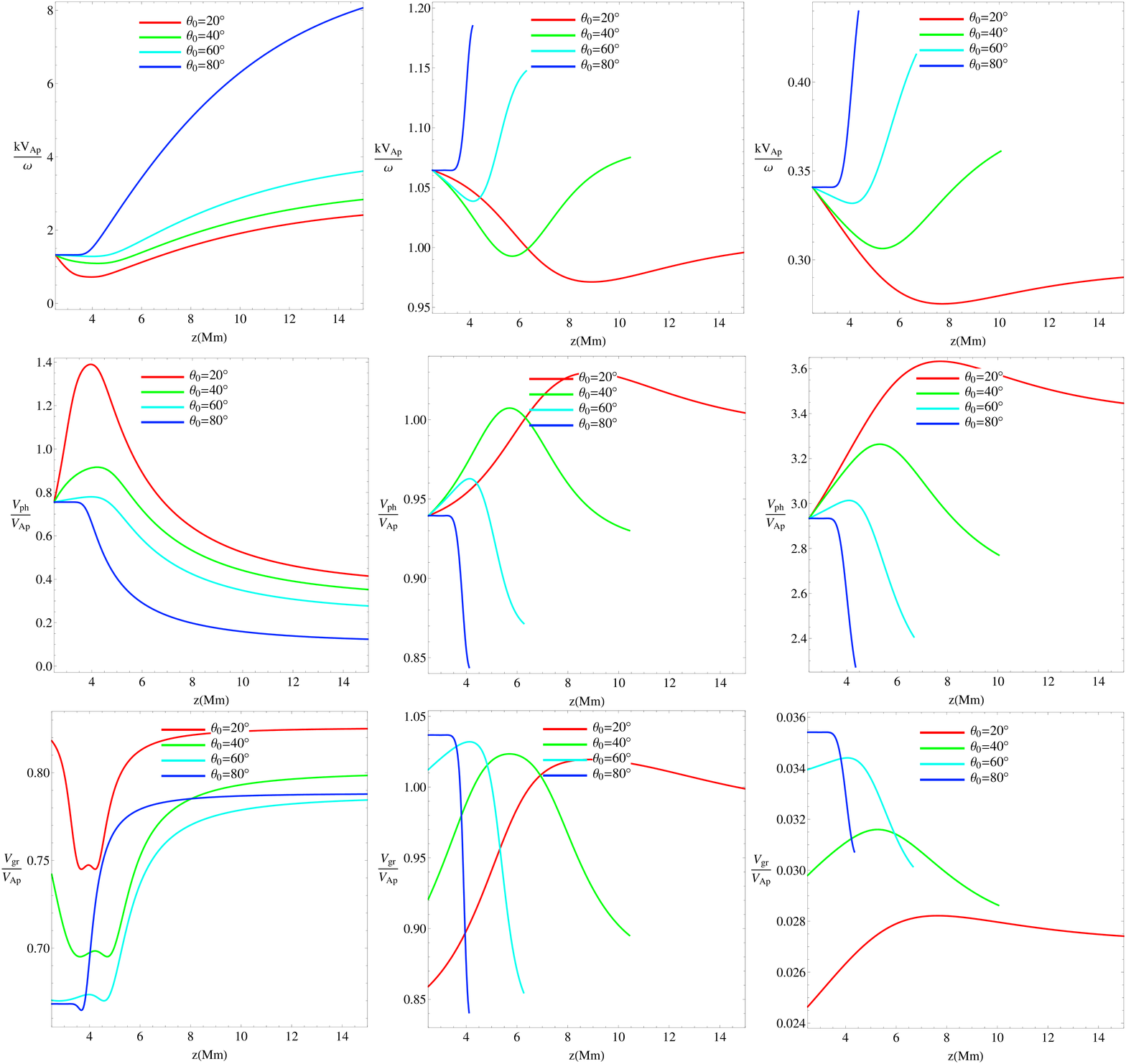}
\end{center}
\vspace{-0.4cm} \caption{Variation of the basic properties of the
modes as a function of height z in the solar atmosphere starting at
the launch location (x$_{0}$=7.5 Mm, z$_{0}$=2.5 Mm) in the coronal
funnel (with a \textbf{B}$_{0}$-inclination angle
$\varphi_{0}\approx79^{\circ}$) with an initial normalized wave
number k$_{0}$=0.2. These results are given for different initial
angles of propagation $\theta_{0}$. The wave properties are : the
refractive index $k\textrm{V}_{Ap}/\omega$, the phase velocity
V$_{ph}$ and the group velocity V$_{gr}$.} \label{fig5}
\end{figure}
\begin{figure}
\begin{center}
$\begin{array} {c@{\hspace{0.in}}c@{\hspace{0in}}c}
\multicolumn{1}{c}{\hspace{-0.3cm}\mbox{\small Ion-Cyclotron mode
2}} & \multicolumn{1}{c}{\hspace{0.8cm}\mbox{\small Ion-Cyclotron
mode 1}} & \multicolumn{1}{c}{\hspace{1.5cm}\mbox{\small Fast mode}}
\end{array}$
\includegraphics[width=12cm]{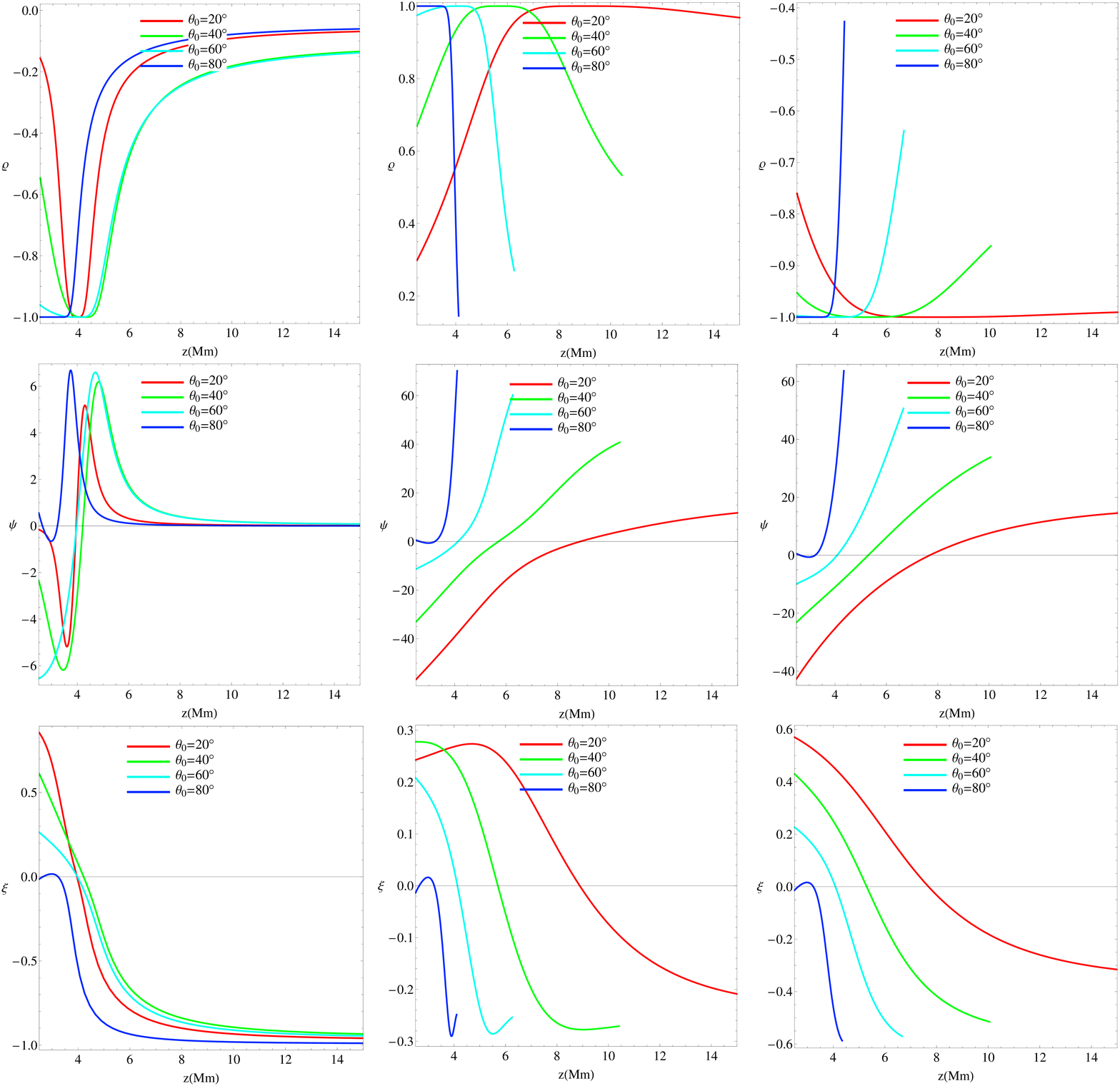}
\end{center}
\vspace{-0.4cm} \caption{Variation of the basic properties of the
modes as a function of height z in the solar atmosphere starting at
the launch location (x$_{0}$=7.5 Mm, z$_{0}$=2.5 Mm) in the coronal
funnel (with a \textbf{B}$_{0}$-inclination angle
$\varphi_{0}\approx79^{\circ}$) with an initial normalized wave
number k$_{0}$=0.2. These results are given for different initial
angles of propagation $\theta_{0}$. The wave properties are : the
helicity (degree of circular polarization) $\varrho$, the angle
between the direction of the group velocity and the ambient magnetic
field $\psi$ and the electrostatic part of the wave $\xi$.}
\label{fig6}
\end{figure}

In this section we go beyond the local treatment of the waves and
perform a non-local wave study using the ray-tracing differential
equations. They are solved employing initial conditions obtained
from the local solutions of the dispersion relation Eq.
(\ref{dispersion}) presented in the previous section. The wave is
launched at the initial location ($x_{0}=7.5$~Mm and $z_{0}=2.5$~Mm)
with different initial angle of propagation,
$\theta_{0}=20^{\circ},40^{\circ},60^{\circ}~\textrm{and}~80^{\circ}$.
The computation have been performed for different initial values of
the normalized wave number k$_{0}$, but for the sake of consistency
the results will only be given for k$_{0}$=0.2. The conclusions
which will be drawn are valid for all waves emitted at
$\textrm{k}_{0}<1$. The ray trajectory of the waves while
propagating in the funnel are shown on the bottom panels of
Fig.(\ref{fig4}) and the spatial variation of their properties as a
function of the $z$-coordinate are presented on Fig.(\ref{fig5}) and
(\ref{fig6}). On the top panels of Fig.(\ref{fig4}) are shown the
polar plots of the group velocity for the three modes corresponding
to different normalized wave numbers
$\textrm{kV}_{Ap}$/$\Omega_{p}$=0.2, 0.4,0.6 and 0.8. They reveal
that the IC2 mode propagation is mainly anisotropic in the direction
of the ambient field \textbf{B}$_{0}$ and cannot propagate
perpendicular to it. Thus the corresponding energy mainly flows
along the magnetic field lines, with maximum group velocity at
parallel propagation. On the other hand, the IC1 mode and the fast
mode propagate almost isotropically with no preferred direction, and
consequently the energy flow is fairly isotropic.\\
These results are confirmed by the ray-tracing computation which
clearly shows that the IC2 mode is well guided along the field
lines, since the ray paths (direction of the group velocity) for
various initial angle of propagation $\theta_{0}$ nicely follow the
magnetic field lines (bottom panels of Fig.\ref{fig4}). Indeed, as
shown on Fig.(\ref{fig6}), the maximal angular deviation is
$\mid\psi\mid\approx6.5^{\circ}$ at the lower part of the funnel
$3<z<5 \textrm{Mm}$ whereas the propagation is quasi-parallel, i.e.
$\psi\approx0^{\circ}$, in its upper part. During this
quasi-parallel propagation the IC2 mode is quasi-electrostatic, i.e.
$\mid\xi\mid\approx1$, and is elliptically polarized with
$0.1<\mid\varrho\mid<0.4$.\\
Contrary, the IC1 mode and the Fast mode are unguided modes with
their ray path having mainly a strait trajectory (bottom panels of
Fig.\ref{fig4}). The angular deviation $\psi$ between the direction
of the group velocity and \textbf{B}$_{0}$, shown on
Fig.(\ref{fig6}), varies between $-50^{\circ}$ and $70^{\circ}$ for
the IC1 mode and $-40^{\circ}$ and $60^{\circ}$ for the Fast mode.
The polarization is in general elliptical, right-handed
($\varrho>0$) for the IC1 mode and left-handed ($\varrho<0$) for the
Fast mode, except for $\theta=20^{\circ}$ in the upper part of the
funnel (z$>=6$ Mm) where it is mainly circular right-handed
($\varrho=+1$) for the IC1 mode and circular left-handed
($\varrho=-1$) for the Fast mode.

\section{Conclusion}\label{Conclusion}

Ion-cyclotron waves propagation in a coronal funnel have been
studied via a linear mode analysis using the cold three-fluid
(e-p-He$^{2+}$) model.\\
First local solutions of the dispersion relation have been given for
a defined region in the funnel. For comparison purpose dispersion
curves in the case of a two-fluid (e-p) model have been also
presented. They showed the presence of two modes which are the
extensions of the standard Alfv\'{e}n and Fast MHD modes into the
high-frequency domain ($\omega\approx\Omega_{p}$) where they are
dispersive. At large wave number, the Alfv\'{e}n mode (also named
Ion-Cyclotron mode) experiences a resonance regime at
$\omega=\Omega_{p}$. In the three-fluid model, the results have
shown that the consideration of a second ion population, namely
alpha particles He$^{2+}$, strongly influences the dispersion
branches as compared to the two-fluid model. Indeed, they are
subject to mode coupling and conversion and show the appearance of a
cut-off frequency (concerning the Fast mode). In addition to the
first Ion-Cyclotron mode "IC1" and the Fast mode previously present
in the two-fluid case, we note the appearance of a second
Ion-Cyclotron mode "IC2" also subject to a resonance regime at
$\omega=\Omega_{\alpha}$ at large k.\\
The nonlocal wave analysis, which has been performed by solving the
ray tracing equations, revealed that the IC2 mode is very well
guided by the magnetic field since its ray path nicely follows the
field lines independently of the initial angle of propagation. This
is in agreement with the topology of the IC2 mode group velocity
which predicted a fairly anisotropic direction of propagation in the
direction of the magnetic field with a maximum value parallel to it.
As a consequence, the energy associated with this mode can
carried-out to the outer corona along the open magnetic field lines.
On the other hand, the IC1 mode and the Fast mode are unguided since
their ray paths do not follow the magnetic field lines. They can
propagate in the funnel isotropically with approximately a strait
trajectory. The basic wave properties in term of phase and group
velocity, polarization and other important quantities show in
general a significant variation as the waves propagate in the
funnel, revealing thus the importance of the non-uniformity of the
medium. All these effects, and particularly the possible wave
absorption and dissipation near the resonance frequencies of minor
heavy ions might play a role in coronal heating.

% uncomment below if you wish to put bibiliography (very highly
% NOT recommended)
% \begin{thebibliography}{}
%\bibitem[]{}
%\end{thebibliography}

\bibliographystyle{bib-sty}
\bibliography{Mecheri-2012}
\end{document}